\documentclass
[showpacs,pra,superscriptaddress,10pt,twocolumn,bibnotes]{revtex4}%
\usepackage{amsfonts}
\usepackage{amsmath}
\usepackage{amssymb}
\usepackage{graphicx}%
\newtheorem{theorem}{Theorem}

\newtheorem{lemma}[theorem]{Lemma}

\newenvironment{proof}[1][Proof]{\noindent\textbf{#1.} }{\ \rule{0.5em}{0.5em}}
\ifx\pdfoutput\relax\let\pdfoutput=\undefined\fi
\newcount\msipdfoutput
\ifx\pdfoutput\undefined\else
\ifcase\pdfoutput\else
\ifx\paperwidth\undefined\else
\ifdim\paperheight=0pt\relax\else\pdfpageheight\paperheight\fi
\ifdim\paperwidth=0pt\relax\else\pdfpagewidth\paperwidth\fi
\fi\fi\fi
\begin{document}
\title{Quantum state redistribution based on a generalized decoupling}
\author{Ming-Yong Ye}
\affiliation{Department of Physics and Center of Theoretical and Computational Physics,
University of Hong Kong, Pokfulam Road, Hong Kong, People's Republic of China}
\affiliation{School of Physics and Optoelectronics Technology, Fujian Normal University,
Fuzhou 350007, People's Republic of China}
\author{Yan-Kui Bai}
\affiliation{Department of Physics and Center of Theoretical and Computational Physics,
University of Hong Kong, Pokfulam Road, Hong Kong, People's Republic of China}
\affiliation{College of Physical Science and Information Engineering, Hebei Normal
University, Shijiazhuang, Hebei 050016, Pepole's Republic of China}
\author{Z. D. Wang}
\affiliation{Department of Physics and Center of Theoretical and Computational Physics,
University of Hong Kong, Pokfulam Road, Hong Kong, People's Republic of China}

\begin{abstract}
We develop a simple protocol for a one-shot version of quantum state
redistribution, which is the most general two-terminal source coding problem. The
protocol is simplified from a combination of protocols for the fully quantum
reverse Shannon and fully quantum Slepian-Wolf problems, with its time-reversal symmetry
being apparent. When the protocol is applied to the case where the
redistributed states have a tensor power structure, more natural
resource rates are obtained.

\end{abstract}

\maketitle

\textit{Introduction}--Quantum information theory may be understood in terms
of interconversion between various resources \cite{privacy,frame}. In this
resource framework, the well-known quantum teleportation \cite{teleport} can
be regarded as a process in the simulation of noiseless quantum channels using
entanglement plus noiseless classical channels. Significant efforts and
progress have been made in the unification of quantum information
theory\cite{conditional,res,sw,family}; many important results in quantum
information theory, such as one-way entanglement distillation \cite{privacy},
state merging \cite{nature}, fully quantum Slepian-Wolf theorem (FQSW)
\cite{sw}, fully quantum reverse Shannon theorem (FQRS) \cite{sw,triangle}, and
quantum channel capacities \cite{c3,c4,c5,c6,c7}, can be derived
in the framework of quantum state redistribution (QSR) \cite{conditional,res}.

\begin{figure}
[ptb]
\begin{center}
\includegraphics[scale=0.40]{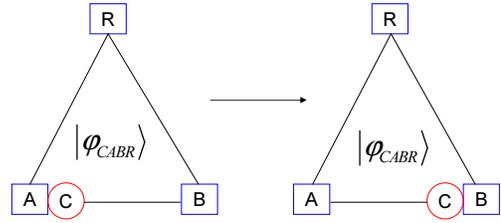}
\caption{(Color online). Quantum state redistribution. Initially $A$ and $C$
is held by Alice, $B$ is held by Bob and $R$ is the reference system. The task
is to redistribute $C$ from Alice to Bob. If Bob's side information $B$ is
viewed as a part of the reference system, it is reduced to FQRS problem; If
Alice's side information $A$ is viewed as a part of the reference system, it
is reduced to FQSW problem.}
\label{fig1}
\end{center}
\end{figure}

The one-shot version of QSR refers to a communication scenario where Alice and
Bob share a quantum state $\varphi_{CAB}$ in which Alice holds $AC$ and Bob
holds $B$. The shared state $\varphi_{CAB}$ can be viewed as the reduced
state of $\left\vert \varphi_{CABR}\right\rangle $ where $R$ is an inaccessible reference system.
The task is to redistribute $C$ to Bob while trying to keep the whole pure
state unchanged [see FIG. \ref{fig1}]. As in
the classical information theory, it is interesting to consider the case where
Alice, Bob and the reference system share many copies of $\left\vert
\varphi_{CABR}\right\rangle $ and the task is to redistribute all $C$ from
Alice to Bob. Since each copy of the state is identical and independently
distributed this is often called the i.i.d. case. To accomplish the task,
protocols are allowed to use noiseless quantum communication and entanglement.
Minor imperfections in the final state are tolerable provided that they vanish
in the asymptotic region, i.e., when the number of copies goes to infinity. The
minimal resources needed per copy in the asymptotic region are important and
useful information-theoretic quantities.

QSR was first studied by Luo and Devetak\cite{luo}, where the necessary
resources needed per copy in the asymptotic region of the i.i.d. case were
given. Later, Yard and Devetak indicated that the necessary resources are also
sufficient~\cite{res}, i.e., the redistribution of each copy can be accomplished by sending
$Q=I\left(  C;R\left\vert B\right.  \right)  /2$ qubits to Bob and consuming
$E=\left[  I(A;C)-I(B;C)\right]  /2$ ebits in the asymptotic region,
here $I\left(  C;R\left\vert B\right.  \right)  $ is quantum conditional
mutual information and $I(A;C)$ is quantum mutual information defined on
$\left\vert \varphi_{CABR}\right\rangle $ . If $E$ is negative, instead of
be consumed,
it is distilled. However in the process of obtaining the resource rates, coherent
channels \cite{coherent} plus the cancelation lemma were used, which leads
the process to be more complicated \cite{res}.

\begin{figure}
[ptb]
\begin{center}
\includegraphics[scale=0.46]{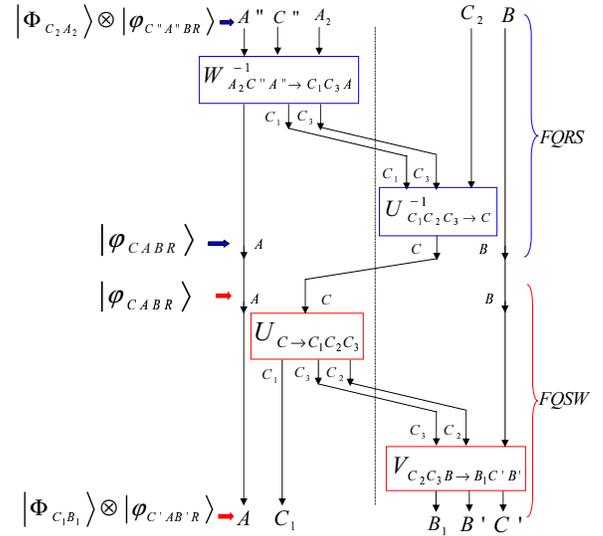}
\caption{(Color online) A non-optimal protocol for one-shot version of QSR
that is a direct combination of protocols for FQRS and FQSW problems. The
decoding operation $U^{-1}$ in FQRS is the inversion of the encoding operation
$U$ in FQSW, so their effect cancels out. }
\label{fig2}
\end{center}
\end{figure}

In this paper, we first develop a simple protocol for a one-shot version of QSR,
which is simplified from a direct combination
of protocols for FQRS and FQSW problems [see FIG. \ref{fig2}]. The
simplification can be done based on a generalized decoupling.
The application of our protocol in the i.i.d. case shows the redistribution of each copy can be accomplished by
sending $Q=I\left(  C;R\left\vert B\right.  \right)  /2$ qubits from Alice to
Bob, consuming $E_{1}=I(A;C)/2$ ebits and distilling $E_{2}=I(B;C)/2$ ebits
at the same time in the asymptotic region. This result looks more natural and the net
entanglement consumed is also $E=E_{1}-E_{2}$ ebits per copy
\cite{conditional,res}.

\textit{QSR: one-shot version}--
We first give some notations. The density operator of state $\left\vert
\varphi\right\rangle $ is denoted by $\varphi$. We use $d_{A}$ to denote the
dimension of $A$ and $U\cdot\rho$ to denote the adjoint action of $U$ on
$\rho$, i.e., $U\cdot\rho=U\rho U^{-1}$. The trace norm of $\rho$ is defined
as $\left\Vert \rho\right\Vert _{1}=Tr\sqrt{\rho^{\dagger}\rho}$.

\begin{lemma}
\label{lem2} (lemma 2.2 of \cite{frame}) For two pure states $\left\vert
\mu_{AB}\right\rangle $ and $\left\vert \nu_{AC}\right\rangle $, if the
corresponding reduced states $\mu_{A}$ and $\nu_{A}$ satisfy $\left\Vert
\mu_{A}-\nu_{A}\right\Vert _{1}\leq\epsilon$, there exists an isometry
$K_{B\rightarrow C}$ such that $\left\Vert K_{B\rightarrow C}\cdot\mu_{AB}%
-\nu_{AC}\right\Vert _{1}\leq2\sqrt{\epsilon}$.
\end{lemma}

\begin{theorem}
\label{th1}(generalized decoupling theorem) For any given density operators
$\omega_{CF}$ and $\psi_{CE}$ in which $C=C_{1}C_{2}C_{3}$, there exists a
unitary operation $U$ on $C$ such that
\begin{align}
\left\Vert Tr_{C_{2}C_{3}}[U\cdot\omega_{CF}]-\pi_{C_{1}}\otimes\omega
_{F}\right\Vert _{1}^{2}  &  \leq2\alpha,\label{s1}\\
\left\Vert Tr_{C_{1}C_{3}}[U\cdot\psi_{CE}]-\pi_{C_{2}}\otimes\psi
_{E}\right\Vert _{1}^{2}  &  \leq2\beta, \label{s2}%
\end{align}
where $\pi_{C_{1}}$ and $\pi_{C_{2}}$ are maximal mixed states, $\omega_{F}$
and $\psi_{E}$ are the reduced states from $\omega_{CF}$ and $\psi_{CE}$. The
upper bounds are defined as $\alpha=d_{C}d_{F}Tr\left(  \omega_{CF}\right)
^{2}\left/  d_{C_{2}C_{3}}^{2}\right.  $ and $\beta=d_{C}d_{E}Tr\left(
\psi_{CE}\right)  ^{2}\left/  d_{C_{1}C_{3}}^{2}\right.  $.
\end{theorem}

\begin{proof}
The decoupling theorem \cite{res,sw,father,capacity} states
\begin{align}%
{\displaystyle\int\limits_{\mathbb{U}\left(  C\right)  }}
\left\Vert Tr_{C_{2}C_{3}}[U\cdot\omega_{CF}]-\pi_{C_{1}}\otimes\omega
_{F}\right\Vert _{1}^{2}dU  &  \leq\alpha,\\%
{\displaystyle\int\limits_{\mathbb{U}\left(  C\right)  }}
\left\Vert Tr_{C_{1}C_{3}}[U\cdot\psi_{CE}]-\pi_{C_{2}}\otimes\psi
_{E}\right\Vert _{1}^{2}dU  &  \leq\beta.
\end{align}
These inequalities indicate more than one half of the unitary operators $U$
on $C$ satisfying (\ref{s1}) and more than one half of the unitary operators
$U$ on $C$ satisfying (\ref{s2}), so there exists a unitary operation $U$ on
$C$ such that both (\ref{s1}) and (\ref{s2}) are satisfied at the same time.
\end{proof}

We introduce two reference states $\left\vert \hat{\varphi}_{CABR}%
\right\rangle $ and $\left\vert \check{\varphi}_{CABR}\right\rangle $ in
discussing the redistribution of $\left\vert \varphi_{CABR}\right\rangle $ and
define
\begin{subequations}
\label{bv}%
\begin{align}
\gamma_{1} &  =2\left\Vert \varphi_{CABR}-\hat{\varphi}_{CABR}\right\Vert
_{1},\\
\gamma_{2} &  =2\left\Vert \varphi_{CABR}-\check{\varphi}_{CABR}\right\Vert
_{1}.
\end{align}
Assume that $C=C_{1}C_{2}C_{3}$, $A^{\prime\prime}$ and $B^{\prime}$ are the
duplicates of $A$ and $B$, $C^{\prime}$ and $C^{\prime\prime}$ are the
duplicates of $C$. The generalized decoupling theorem \ref{th1} ensures that
we are able to approximately decouple $C_{2}$ from $BR$ in $\left\vert \hat{\varphi
}_{CABR}\right\rangle $ and approximately decouple $C_{1}$ from $AR$ in $\left\vert
\check{\varphi}_{CABR}\right\rangle $ by the same operation on $C$. Combining
with the lemma \ref{lem2} and following the similar deduction in \cite{sw}, it
can be seen that there exist the unitary operation $U$ on $C$, the isometries
$W_{C_{1}C_{3}A\rightarrow A_{2}C^{\prime\prime}A^{\prime\prime}}$, and
$V_{C_{2}C_{3}B\rightarrow B_{1}C^{\prime}B^{\prime}}$ such that%
\end{subequations}
\begin{align}
\left\Vert \left(  W\circ U\right)  \cdot\hat{\varphi}_{CABR}-\Phi_{C_{2}%
A_{2}}\otimes\hat{\varphi}_{C^{\prime\prime}A^{\prime\prime}BR}\right\Vert
_{1} &  \leq\eta_{1},\label{s}\\
\left\Vert \left(  V\circ U\right)  \cdot\check{\varphi}_{CABR}-\Phi
_{C_{1}B_{1}}\otimes\check{\varphi}_{C^{\prime}AB^{\prime}R}\right\Vert _{1}
&  \leq\eta_{2},\label{ss}%
\end{align}
where $\Phi_{C_{2}A_{2}}$ and $\Phi_{C_{1}B_{1}}$ are the maximally entangled
pure states, $\hat{\varphi}_{C^{\prime\prime}A^{\prime\prime}BR}$ and
$\check{\varphi}_{C^{\prime}AB^{\prime}R}$ are the same pure states as
$\hat{\varphi}_{CABR}$ and $\check{\varphi}_{CABR}$ respectively. The upper
bounds $\eta_{1}$ and $\eta_{2}$ are defined as
\begin{subequations}
\label{vb}%
\begin{align}
\eta_{1} &  =2\sqrt[4]{2d_{C}d_{BR}Tr\left(  \hat{\varphi}_{CBR}\right)
^{2}\left/  d_{C_{1}C_{3}}^{2}\right.  },\\
\eta_{2} &  =2\sqrt[4]{2d_{C}d_{AR}Tr\left(  \check{\varphi}_{CAR}\right)
^{2}\left/  d_{C_{2}C_{3}}^{2}\right.  }.
\end{align}
Making use of the triangle inequality, we have from (\ref{s}) and (\ref{ss})
\end{subequations}
\begin{align}
\left\Vert \left(  W\circ U\right)  \cdot\varphi_{CABR}-\Phi_{C_{2}A_{2}%
}\otimes\varphi_{C^{\prime\prime}A^{\prime\prime}BR}\right\Vert _{1} &
\leq\Delta_{1},\label{f11}\\
\left\Vert \left(  V\circ U\right)  \cdot\varphi_{CABR}-\Phi_{C_{1}B_{1}%
}\otimes\varphi_{C^{\prime}AB^{\prime}R}\right\Vert _{1} &  \leq\Delta
_{2},\label{f12}%
\end{align}
where $\Delta_{i}=\gamma_{i}+\eta_{i}$, $\varphi_{C^{\prime\prime}%
A^{\prime\prime}BR}$ and $\varphi_{C^{\prime}AB^{\prime}R}$ are the same pure
state as $\varphi_{CABR}$.%

In Fig.~\ref{fig2} we depict a protocol for redistributing $\left\vert
\varphi_{C^{\prime\prime}A^{\prime\prime}BR}\right\rangle $ where Alice holds
$C^{\prime\prime}A^{\prime\prime}$ and Bob holds $B$, i.e., redistributing
$C^{\prime\prime}$ from Alice to Bob. This protocol is non-optimal since it
cannot achieve the optimal resource rates in the i.i.d. case. However, its
simplified version shown in Fig.~\ref{fig3} is optimal. The non-optimal
protocol in Fig.~\ref{fig2} consists of three steps. The first step is to
redistribute $C^{\prime\prime}$ from Alice to Bob via the protocol for FQRS
problem \cite{sw} where Bob's side information $B$ is treated as a part of the
reference system. In this step, the maximally entangled state $\Phi
_{C_{2}A_{2}}$ is consumed and the system $C_{1}C_{3}$ is transmitted from
Alice to Bob via noiseless quantum channels. According to (\ref{f11}) the
system is in $\left\vert \varphi_{CABR}\right\rangle $ after completing this
step if $\Delta_{1}=0$. The second step is to send $C=C_{1}C_{2}C_{3}$ from
Bob back to Alice via noiseless quantum channels. The third step is the
process that redistributes $C$ from Alice to Bob via the protocol for FQSW
problem \cite{sw} where Alice's side information $A$ is treated as a part of
the reference system. In this step the system $C_{2}C_{3}$ is transmitted from
Alice to Bob via noiseless quantum channels. According to (\ref{f12}) the
system is in $\left\vert \Phi_{C_{1}B_{1}}\right\rangle \otimes\left\vert
\varphi_{C^{\prime}AB^{\prime}R}\right\rangle $ after this step if $\Delta
_{1}=\Delta_{2}=0$, i.e., the maximally entangled state $\Phi_{C_{1}B_{1}}$ is
distilled and the redistribution is perfectly accomplished since $\left\vert
\varphi_{C^{\prime}AB^{\prime}R}\right\rangle $ is the same state as
$\left\vert \varphi_{C^{\prime\prime}A^{\prime\prime}BR}\right\rangle $ but
with $A$ on Alice's side and $C^{\prime}B^{\prime}$ on Bob's side.

For the above described non-optimal protocol, the decoding operation
$U_{C_{1}C_{2}C_{3}\rightarrow C}^{-1}$ in the first step is the inversion of
the encoding operation $U_{C\rightarrow C_{1}C_{2}C_{3}}$ in the third step,
which is the result of the generalized decoupling theorem. So their effects
cancel out and the whole process can be simplified. Our protocol for
redistributing $C^{\prime\prime}$ in $\left\vert \varphi_{C^{\prime\prime
}A^{\prime\prime}BR}\right\rangle $ from Alice to Bob is just the simplified
version of the above one and it consists of three steps [see FIG. \ref{fig3}]:

\begin{enumerate}
\item Alice first implements the encoding operation $W_{A_{2}C^{\prime\prime
}A^{\prime\prime}\rightarrow C_{1}C_{3}A}^{-1}$ on the initial pure state
$\Phi_{C_{2}A_{2}}\otimes\varphi_{C^{\prime\prime}A^{\prime\prime}BR}$ where
Alice holds $A_{2}C^{\prime\prime}A^{\prime\prime}$ and Bob holds $C_{2}B$.

\item Alice transmits $C_{3}$ to Bob via noiseless quantum channels.

\item Bob makes the decoding operation $V_{C_{2}C_{3}B\rightarrow
B_{1}C^{\prime}B^{\prime}}$.
\end{enumerate}

After these steps, Alice holds $C_{1}A$ and Bob has $B_{1}C^{\prime}B^{\prime
}$. If $\Delta_{1}+\Delta_{2}$ is zero, the final state is exactly the pure
state $\Phi_{C_{1}B_{1}}\otimes\varphi_{C^{\prime}AB^{\prime}R}$; the
redistribution is perfectly accomplished since $\varphi_{C^{\prime}AB^{\prime
}R}$ is the same pure state as $\varphi_{C^{\prime\prime}A^{\prime\prime}BR}$
but with $A$ on Alice's side and $C^{\prime}B^{\prime}$ on Bob's side. In the
process, $\log_{2}d_{C_{3}}$ qubits are transmitted from Alice to Bob,
$\log_{2}d_{C_{2}}$ ebits are consumed and $\log_{2}d_{C_{1}}$ ebits are
distilled. Generally, the distance between the end state $\left(  V\circ
W^{-1}\right)  \cdot\left(  \Phi_{C_{2}A_{2}}\otimes\varphi_{C^{\prime\prime
}A^{\prime\prime}BR}\right)  $ and the pure state $\Phi_{C_{1}B_{1}}%
\otimes\varphi_{C^{\prime}AB^{\prime}R}$ in terms of the trace norm is not longer
than $\Delta_{1}+\Delta_{2}$, which can be derived from (\ref{f11}) and
(\ref{f12}).

\textit{QSR: i.i.d. case}--At this stage, we apply our protocol to the i.i.d.
case and show that $\Delta_{1}+\Delta_{2}$ approaches zero in the
asymptotic region. Our technical result is summarized in a theorem:

\begin{figure}
[ptb]
\begin{center}
\includegraphics[scale=0.65]{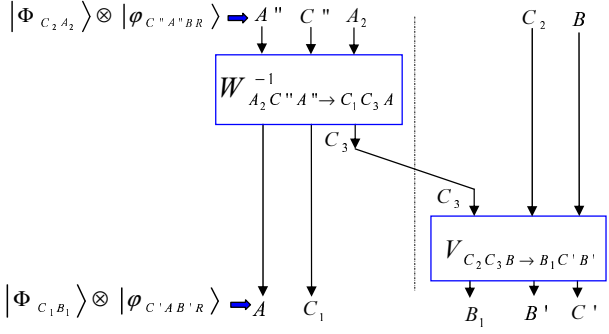}
\caption{(Color online) Our protocol for one-shot version of QSR. The encoding
operation by Alice is $W^{-1}$ and the decoding operation by Bob is $V$. The
reverse redistribution uses $V^{-1}$ and $W$ as the encoding and decoding
operations. The time-reversal symmetry of the protocol is apparent. The states
on the left are obtained if $\Delta_{1}+\Delta_{2}$ is zero.}
\label{fig3}
\end{center}
\end{figure}

\begin{theorem}
\label{th2}(one-shot version of QSR) Suppose that quantum system
$A^{\prime\prime}$ is a duplicate of $A$, $B^{\prime}$ is a duplicate of $B$,
$C^{\prime}$ and $C^{\prime\prime}$ are duplicates of $C$, and $C=C_{1}%
C_{2}C_{3}$. Density operators $\Phi_{C_{1}B_{1}}$ and $\Phi_{C_{2}A_{2}}$
represent maximally entangled pure states. For any given pure states
$\Upsilon_{initial}=\Phi_{C_{2}A_{2}}\otimes\varphi_{C^{\prime\prime}%
A^{\prime\prime}BR}$ and $\Upsilon_{final}=\Phi_{C_{1}B_{1}}\otimes
\varphi_{C^{\prime}AB^{\prime}R}$, there are isometries $V_{C_{2}%
C_{3}B\rightarrow B_{1}C^{\prime}B^{\prime}}$ and $W_{C_{1}C_{3}A\rightarrow
A_{2}C^{\prime\prime}A^{\prime\prime}}$ such that%
\begin{align}
\left\Vert \left(  V\circ W^{-1}\right)  \cdot\Upsilon_{initial}%
-\Upsilon_{final}\right\Vert _{1}  &  \leq\Delta_{1}+\Delta_{2},\label{xv2}\\
\left\Vert \left(  W\circ V^{-1}\right)  \cdot\Upsilon_{final}-\Upsilon
_{initial}\right\Vert _{1}  &  \leq\Delta_{1}+\Delta_{2}, \label{x11}%
\end{align}
where $\varphi_{C^{\prime\prime}A^{\prime\prime}BR}$ and $\varphi_{C^{\prime
}AB^{\prime}R}$ are the same pure state as $\varphi_{CABR}$, the upper bound
$\Delta_{1}+\Delta_{2}$ equals $\gamma_{1}+\eta_{1}+\gamma_{2}+\eta_{2}$ that
can be calculated from Eqs. (\ref{bv}) and (\ref{vb}).
\end{theorem}

Consider the i.i.d. case: Alice, Bob, and the reference system share $n$
copies of the state $\left\vert \varphi_{CABR}\right\rangle $, i.e.,
$\left\vert \Psi_{C^{n}A^{n}B^{n}R^{n}}\right\rangle =\left\vert
\varphi_{CABR}\right\rangle ^{\otimes n}$ with $C^{n}A^{n}$ on Alice's side
and $B^{n}$ on Bob's side, where we use the notion $A^{n}=A_{1}A_{2}\cdots
A_{n}$. The task is to redistribute $C^{n}$ from Alice to Bob. This is the
redistribution of quantum states that have a tensor power structure.

We first recall the relevant knowledge about typical subspace \cite{res,sw}.
Based on quantum state $\left\vert \Psi_{C^{n}A^{n}B^{n}R^{n}}\right\rangle $
we can define $\delta$-typical subspace of $F^{n}$, $F=A,B,C,AR,BR$, which is
the subspace spanned by the eigenstates of $\left(  \varphi_{F}\right)
^{\otimes n}$ with the corresponding eigenvalues $\lambda$ satisfying
$2^{-n\left(  S\left(  F\right)  +\delta\right)  }\leq\lambda\leq2^{-n\left(
S\left(  F\right)  -\delta\right)  }$, where $\delta$ is a small positive
number and $S\left(  F\right)  $ is von Neumann entropy of $\varphi_{F}$. The
projector onto the $\delta$-typical subspace of $F^{n}$ is denoted by
$\Pi_{F^{n}}^{\delta}$. We introduce the three new normalized states:%
\begin{align}
\left\vert \Omega_{C^{typ}A^{n}B^{n}R^{n}}\right\rangle  &  \propto\Pi_{C^{n}%
}^{\delta}\left\vert \Psi_{C^{n}A^{n}B^{n}R^{n}}\right\rangle \\
\left\vert \check{\Omega}_{C^{typ}B^{n}\left(  AR\right)  ^{n}}\right\rangle
&  \propto\Pi_{\left(  AR\right)  ^{n}}^{\delta}\Pi_{B^{n}}^{\delta}\Pi
_{C^{n}}^{\delta}\left\vert \Psi_{C^{n}A^{n}B^{n}R^{n}}\right\rangle \\
\left\vert \hat{\Omega}_{C^{typ}A^{n}\left(  BR\right)  ^{n}}\right\rangle  &
\propto\Pi_{\left(  BR\right)  ^{n}}^{\delta}\Pi_{A^{n}}^{\delta}\Pi_{C^{n}%
}^{\delta}\left\vert \Psi_{C^{n}A^{n}B^{n}R^{n}}\right\rangle
\end{align}
where $C^{typ}$ is used to denote the support of $\Pi_{C^{n}}^{\delta}$. For a
sufficiently large $n$, we have%
\begin{equation}
\left\Vert \Psi_{C^{n}A^{n}B^{n}R^{n}}-\Omega_{C^{typ}A^{n}B^{n}R^{n}%
}\right\Vert _{1}\leq e^{-c\delta^{2}n} \label{w3}%
\end{equation}
with a positive constant number $c$.

To redistribute $C^{n}$ in $\left\vert \Psi_{C^{n}A^{n}B^{n}R^{n}}\right\rangle$
 from Alice to Bob, Alice first makes a projective
measurement on $C^{n}$ with projectors $\Pi_{C^{n}}^{\delta}$ and $I_{C^{n}%
}-\Pi_{C^{n}}^{\delta}$. If the outcome of $I_{C^{n}}-\Pi_{C^{n}}^{\delta}$ is
obtained, then the protocol fails. If the outcome of $\Pi_{C^{n}}^{\delta}$ is
obtained, then the systems is in the state $\left\vert \Omega_{C^{typ}%
A^{n}B^{n}R^{n}}\right\rangle $. The inequality (\ref{w3}) indicates that the
probability of failure vanishes when $n$ approaches to infinite. When quantum
state $\Omega_{C^{typ}A^{n}B^{n}R^{n}}$ is obtained, they then redistribute
$C^{typ}$ from Alice to Bob via our protocol for one-shot version of QSR with
the reference states $\hat{\Omega}_{C^{typ}A^{n}\left(  BR\right)  ^{n}}$ and
$\check{\Omega}_{C^{typ}B^{n}\left(  AR\right)  ^{n}}$.

We use $\Delta_{1}+\Delta_{2}$ to describe how well the redistribution is
accomplished. A key point is that dimensions and purities in the expression of
$\Delta_{1}+\Delta_{2}$ now have bounds that are related to entropy quantities
\cite{res,sw}. Assume that $n$ is large enough and $C^{typ}=C_{1}^{n}C_{2}%
^{n}C_{3}^{n}$ with dimensions%
\begin{align}
\log_{2}d_{C_{1}^{n}}  &  =n\left[  I\left(  B;C\right)  -6t\delta\right]
/2,\\
\log_{2}d_{C_{2}^{n}}  &  =n\left[  I\left(  A;C\right)  -6t\delta\right]
/2,\\
\log_{2}d_{C_{3}^{n}}  &  =n\left[  I\left(  C;R\left\vert B\right.  \right)
+12t\delta+2\eta\right]  /2,
\end{align}
for some $\eta\in\left[  -t\delta,t\delta\right]  $, where $t$ is a constant
bigger than one, $I\left(  B;C\right)  $ and $I\left(  C;R\left\vert B\right.
\right)  $ are quantum mutual information and conditional mutual information
defined on $\left\vert \varphi_{CABR}\right\rangle $ \cite{nielsen}. The exact
value of $\eta$ is determined by the dimension of $C^{typ}$. It can be shown
for a large $n$
\begin{equation}
\Delta_{1}+\Delta_{2}\leq4\left(  e^{-c\delta^{2}n}+\sqrt[4]{2\times
2^{-n\left(  2\eta+3t\delta\right)  }}\right)  ,
\end{equation}
which goes to zero in the asymptotic region.

Note that $\Omega_{C^{typ}A^{n}B^{n}R^{n}}$ and $\Psi_{C^{n}A^{n}B^{n}R^{n}}$
have no difference when $n$ goes to infinity [see (\ref{w3})] and $\delta$ can
be arbitrarily small, we can say that the redistribution of each copy of
$\left\vert \varphi_{CABR}\right\rangle $
can be accomplished by sending
$Q=I\left(  C;R\left\vert B\right.  \right)  /2$ qubits from Alice to Bob,
consuming $E_{1}=I(A;C)/2$ ebits and distilling $E_{2}=I(B;C)/2$ ebits
in the asymptotic region. These are the optimal rates as shown in
\cite{luo}. The reverse redistribution process can be accomplished by sending
$Q$ qubits from Bob to Alice, consuming $E_{2}$ ebits and distilling $E_{1}$
ebits per copy in the asymptotic region. The contents of these two processes
can be expressed via a simple formula%
\begin{equation}
\varphi_{AC|B}+E_{1}\left[  qq\right]  \overset{Q}{\iff}\varphi_{A|CB}%
+E_{2}\left[  qq\right]  , \label{d}%
\end{equation}
where $\varphi_{AC|B}$ represents the quantum state $\varphi_{CAB}$ with $AC$
on Alice's side and $B$ on Bob's side, $\varphi_{A|CB}$ represents the same
quantum state $\varphi_{CAB}$ but with $A$ on Alice's side and $BC$ on Bob's
side, and $\left[  qq\right]  $ denotes an ebit between Alice and Bob
\cite{res}. We emphasize that the formula should be understood in the
asymptotic region of the i.i.d. case. Also notably, $E_{1}$ and $E_{2}$ change
into each other under the exchange of $A$ and $B$, and $Q$ is invariant under
this exchange, namely, (\ref{d}) is symmetric.

\textit{Summary}--We have developed a protocol for one-shot version of QSR,
which can lead to more natural resource rates in the i.i.d. case. The protocol
is simplified from a combination of protocols for FQRS and FQSW problems. The
simplification is based on a generalized decoupling theorem, which may have
further applications in more complicated communication cases such as quantum
state exchange \cite{exchange} and multipartite communication. Since FQRS
problem is the reverse of FQSW problem \cite{sw}, the time-reversal symmetry
is apparent in our protocol.

\textit{Note added}--Recently Oppenheim
studied the QSR from the viewpoint of coherent state-merging \cite{another}.
Our protocol presented here removes the decoupling operations, giving it the
advantage of simplicity which makes it more promising to explore
redistributing quantum states with structures other than the tensor power in
the i.i.d. case.

The authors would like to thank Fei Ye for many useful discussions. The work
was supported by the RGC of Hong Kong under HKU7051/06P, HKU7044/08P, and
HKU-3/05C, and NSF-China Grant No. 10429401. M. Y. Ye was also supported by
the Foundation for Universities in Fujian Province (Grant No. 2007F5041) and NSF-China Grant No. 60878059.


\begin{thebibliography}{99}                                                                                               %


\bibitem {privacy}I. Devetak and A. Winter, Phys. Rev. Lett. \textbf{93},
080501 (2004).

\bibitem {frame}I. Devetak, A.W. Harrow, A. Winter, quant-ph/0512015.

\bibitem {teleport}C. H. Bennett, G. Brassard, C. Cr\'{e}peau, R. Jozsa, A.
Peres and W. K. Wootters, Phys. Rev. Lett. \textbf{70}, 1895 (1993).

\bibitem {conditional}I. Devetak and J. Yard, Phys. Rev. Lett. \textbf{100},
230501 (2008).

\bibitem {res}J. Yard and I. Devetak, arXiv:0706.2907.

\bibitem {sw}A. Abeyesinghe, I. Devetak, P. Hayden, and A. Winter, quant-ph/0606225.

\bibitem {family}I. Devetak, A. W. Harrow, and A. Winter, Phys. Rev. Lett.
\textbf{93}, 230504 (2004).

\bibitem {nature}M. Horodecki, J. Oppenheim and A. Winter, Nature,
\textbf{436}, 673 (2005).

\bibitem {triangle}I. Devetak, Phys. Rev. Lett. \textbf{97}, 140503 (2006).

\bibitem {c3}C. H. Bennett, P. W. Shor, J. A. Smolin, and A. V. Thapliyal,
Phys. Rev. Lett. \textbf{83}, 3081 (1999).

\bibitem {c4}C. H. Bennett, P. W. Shor, J. A. Smolin, and A. V. Thapliyal,
IEEE Trans. Inf. Theory \textbf{48}, 2637 (2002).

\bibitem {c5}S. Lloyd, Phys. Rev. A \textbf{55}, 1613 (1997).

\bibitem {c6}P. W. Shor, www.msri.org/publications/ln/msri/2002 /quantumcrypto/shor/1/.

\bibitem {c7}I. Devetak, IEEE Trans. Inf. Theory \textbf{51}, 44 (2005).

\bibitem {luo}Z. Luo and I. Devetak, quant-ph/0611008.

\bibitem {coherent}A. Harrow, Phys. Rev. Lett. \textbf{92}, 097902 (2004).

\bibitem {father}F. Dupuis and P. Hayden, quant-ph/0612155.

\bibitem {capacity}P. Hayden, M. Horodecki, J. Yard, and A. Winter, Open Syst.
Inf. Dyn. \textbf{15}, 7 (2008).

\bibitem {nielsen}M. A. Nielsen and I. L. Chuang, \textit{Quantum Computation
and Quantum Information} (Cambridge University Press, Cambridge, England, 2000).

\bibitem {exchange}J. Oppenheim, and A. Winter, quant-ph/0511082.

\bibitem {another}J. Oppenheim, arXiv:0805.1065.
\end{thebibliography}
\end{document}